\documentclass[11pt]{article}
\usepackage{longtable}

\usepackage{amsmath}

\usepackage{amssymb}

\begin{document}




\title{{Table of hyperfine anomaly in atomic systems}}

  \author{J.R. Persson\\
  Department of Physics\\
  NTNU\\
  NO-7491 Trondheim\\
  Norway\\
  E-mail: jonas.persson@ntnu.no}



\maketitle

\begin{abstract}
 This table is a compilation of experimental values of magnetic hyperfine anomaly in atomic and ionic systems. The last extensive compilation was published in 1984 by Buttgenbach [Hyperfine Interactions \textbf{20} (1984) 1] and the aim here is to make an up to date compilation. The literature search covers the period to January 2011.
\end{abstract}





\newpage



\section{Introduction}

The atomic electron-nuclear hyperfine interactions have been used to obtain nuclear spins and nuclear multipole moments \cite{otten,billowes}. Isotopic shifts in spectral lines allows determination of the variation of the distribution of nuclear charge, essentially $\Delta\langle r^2 \rangle$ . Experiments on the magnetic counterpart, the distribution of nuclear magnetisation, are more difficult and only a few systematic measurements have been performed \cite{stroke,mosko,Buttgenbach}.The effect of an extended nuclear magnetisation is manifested by the difference between the point-like and actual magnetic dipole hyperfine interaction. This effect was first anticipated by Kopfermann \cite{Kopf} and thought to be too small to be observed. Following the experimental observation by Bitter \cite{Bitter} the effect of extended magnetisation was calculated by Bohr and Weisskopf \cite{bohrweisskopf} in 1950 and is therefore known as the Bohr-Weisskopf (B-W) effect. In order to probe the structural properties of the nucleus, the electron wave function must have a nonzero probability to be found at the origin. This means that only $s_{1/2}$ and relativistic $p_{1/2}$-electrons can be used. Note that electron-electron interaction can cause an s- or p- electron like behaviour, thus giving rise to an apparent nonzero probability for other electrons.

The magnetic dipole hyperfine interaction for the electron-nuclear system is represented by the Hamiltonian
	\begin{equation}
    H=aI \cdot J,
    \end{equation}
Where $a$ is the magnetic dipole hyperfine interaction constant. I and J are the nuclear and electron angular momenta. For an extended nucleus, the point-like hyperfine interaction constant $a_{point}$, is modified by two effects:\\
1.	The modification of the electron wavefunctions by the extended nuclear charge distribution, the "Breit-Rosenthal-Crawford-Schawlow" correction ($\epsilon_{BR}$)\cite{Breit,Crawford,Pallas,Rosenberg}.\\
2.	The extended nuclear magnetisation, the Bohr-Weisskopf effect,($\epsilon_{BW}$ )\cite{bohrweisskopf}.\\
We thus have
    \begin{equation}
    a=a_{point}\left(  1+\epsilon_{BW}\right)\left(  1+\epsilon_{BR}\right)
    \end{equation}
Where $a$ denotes the experimental value of the magnetic dipole hyperfine interaction constant.
The hypothetical $a_{point}$ can not be calculated with sufficient precision for ordinary atoms, as is the case in muonic atoms and hydrogen-like ions. However, these uncertainties in point-like interactions cancel if we take the ratio of the $a$ values for two isotopes:
	\begin{equation}
    \frac{a_{1}}{a_{2}}=
    {\frac{g_I{(1)}}{g_I{(2)}}
    {\frac{[1+\epsilon_{BW}(1)][1+\epsilon_{BR}(1)]}{[1+\epsilon_{BW}(2)][1+\epsilon_{BR}(2)]}}}
    \end{equation}

Where ${g_I = -\mu_I/I}$ is the nuclear gyromagnetic ratio. As the $\epsilon$  are generally small, we get
    \begin{eqnarray}
    \frac{a_{1}}{a_{2}}\approx{\frac{g_I{(1)}}{g_I{(2)}}{[1+\epsilon_{BW}(1)-\epsilon_{BW}(2)][1+\epsilon_{BR}(1)-\epsilon_{BR}(2)]}}\nonumber\\ ={\frac{g_I{(1)}}{g_I{(2)}}{[1+^{1}\Delta^{2}_{BW}][1+^{1}\Delta^{2}_{BR}]}}
    \end{eqnarray}
	
Where we use the definitions for the differential hyperfine anomaly of Bohr-Weisskopf and Breit-Rosenthal corrections, respectively
    \begin{equation}
    ^{1}\Delta^{2}_{BW}\equiv \epsilon_{BW}(1)-\epsilon_{BW}(2)
    \end{equation}
    \begin{equation}
    ^{1}\Delta^{2}_{BR}\equiv \epsilon_{BR}(1)-\epsilon_{BR}(2)
    \end{equation}
	
Calculations of $\epsilon_{BR} $ show significant values, while the differential $^{1}\Delta^{2}_{BR}$ is expected to be small and negligible compared to $^{1}\Delta^{2}_{BW}$\cite{Rosenberg}. However, in cases where the nuclei are very similar  $^{1}\Delta^{2}_{BR}$ will dominate. Dropping the subscripts we obtain
    \begin{equation}
    \frac{a_{1}}{a_{2}}\approx{\frac{g_I{(1)}}{g_I{(2)}}{[1+^{1}\Delta^{2}]}}
    \end{equation}
	
Because the hyperfine anomaly is a quantity of the order of $10^{-3}$ , it is necessary to know the hyperfine interaction constants, $a$, and the nuclear gyromagnetic values with at least an accuracy of $10^{-4}$ or better to obtain values accurate to 10\% for the hyperfine anomaly\cite{Buttgenbach}. Precision values of the hyperfine interaction constants, $a$, and independently measured gyromagnetic ratios are thus needed to obtain the differential hyperfine anomaly,$^{1}\Delta^{2}$.

\section{State-dependent hyperfine anomaly and s-electron hyperfine anomaly}
The hyperfine anomaly shows a state dependence, where the values for different states can vary significantly, but shows an n-independence, as found in Rb \cite{Galvan}. While the hyperfine anomalies normally are on the order of 1\%, state-dependent hyperfine anomaly can attain values up to 10 \%.

The hyperfine interaction can be represented by the following operators\cite{Ref4,Ref5}:
\begin{equation}
h = \frac{{\mu _0 }}{{4\pi }}2\mu _B \sum\limits_{i = 1}^N {\left[ {{\mathop{\rm l}\nolimits} \left\langle {r^{ - 3} } \right\rangle ^{01}  - \sqrt {10} \left( {{\mathop{\rm sC^2}\nolimits}  } \right)^1 \left\langle {r^{ - 3} } \right\rangle ^{12}  + {\mathop{\rm s}\nolimits} \left\langle {r^{ - 3} } \right\rangle ^{10} } \right]} _i  \cdot \mu _I,
\end{equation}
where ${{\mathop{\rm l}\nolimits}}$ and ${\mathop{\rm s}\nolimits}$ are the orbital and spin angularmomentum operators, respectively, of the electron, ${{\mathop{\rm sC^2}\nolimits}  }$ is a tensor product of ${\mathop{\rm s}\nolimits}$ and ${\mathop{\rm C^2}\nolimits}$ of rank 1. The indices stand for the rank in the spin and orbital spaces, respectively. Thus the hyperfine interaction can be considered to consist of three parts, orbital, spin-dipole and contact (spin) interaction, where only the contact (spin) interaction contributes to the hyperfine anomaly . This means that only s and $p_{1/2}$ electrons contribute to the hyperfine anomaly through the contact (spin) interaction. It is, therefore, suitable to rewrite the $a$ constant as
\begin{equation}
a=a_{nc}+a_{c},
\end{equation}

Where $a_{c}$ is the contribution due to the contact interaction and $a_{nc}$ the contribution due to non-contact interactions. The experimental hyperfine anomaly, defined with the experimental $a$ constant, should then be rewritten to obtain the contact contribution to the hyperfine anomaly:
\begin{equation}
{^{1}\Delta_{exp}^{2}}={^{1}\Delta_{c}^{2}}{\frac{a_{c}}{a}}
\end{equation}
where $^{1}\Delta^{2}_{c}$ is the hyperfine anomaly due to the contact interaction, that is, for an $s$- or $p_{1/2}$-electron. The hyperfine anomaly is most often given as the state-dependent hyperfine anomaly, as the s-electron anomaly can be difficult to extract.

Using this result it is possible to extract the anomaly solely from the $a$-constants of two different atomic levels, provided the ratio $a_c/a$ differs substantially for the levels. Comparing the ratio of $a$-constants for two isotopes in two atomic levels, gives:
\begin{equation}
{\frac{{a_{B}^{(1)}/a_{B}^{(2)}}}{{a_{C}^{(1)}/a_{C}^{(2)}}}}\approx
{1+{^{1}\Delta_{c}^{2}}({{\frac{a_{c}^{B}}{a^{B}}}-{\frac{a_{c}^{C}}{{a^{C}}}
})}}
\end{equation}
Where B and C denote different atomic levels and 1 and 2 denote different isotopes. The ratio between the two $a$-constant ratios for the isotopes will only depend on the  difference of the contact contributions of the two atomic levels and the hyperfine anomaly for the s-electron. It should be pointed out that the atomic states used must differ significantly in the ratio $\frac{a_{c}}{a}$, as a small difference will lead to an increased sensitivity to errors \cite{persson}. Considering the special case where ${\frac{a_{c}^{C}}{{a^{C}}}}= 0$, that is when the atomic level does not have any hyperfine anomaly, one can obtain values of the s-electron hyperfine anomaly for level B. This is common practice, however if the state does not exhibit any hyperfine anomaly the hyperfine structure is usually rather small and thus the relative error larger, leading to a large error for the hyperfine anomaly. The optimal case would be two atomic levels within the same multiplet where the experimental (or theoretical) $g_J$ is greater and smaller than 1, respectively.
This is especially useful for unstable isotopes where there high precision measurements of the nuclear magnetic moment do not exist. Furthermore, states with a substantial difference in  the ratio $a_c/a$ are also preferable for studies of the isotope shift. If measurements are performed on more than three atomic levels, it is also possible to deduce the nuclear magnetic moment ratio without the hyperfine anomaly. This will give the nuclear magnetic moment with high accuracy provided the nuclear magnetic moment is known for at least one stable isotope.

\section{Policies followed in the compilation}

The hyperfine anomaly is given with the lightest stable isotope as the reference isotope. The lightest naturally abundant isotope was used for U and the designation of the original article was used for Fr. In most cases the original article, where the hyperfine anomaly has been derived, is used. In the case of recent, more precise values of the nuclear magnetic moment, the hyperfine anomaly has been updated, accordingly. The nuclear magnetic moments of Stone \cite{stone} have been used, unless more precise values of ratios are available. Special care was taken to use magnetic moments obtained by the same method.
The hyperfine anomaly is given as state-dependent if not stated otherwise. If the s-electron hyperfine anomaly is known, no extensive listing of state-dependent hyperfine anomaly is given, unless these are of special interest.




\clearpage

\section*{Table 1.\label{tbl1te} Experimental data of hyperfine anomaly values in atomic systems}

\begin{center}
\begin{tabular}{ll}

Element & The element studied\\

Isotope 1
	& Reference isotope for the hyperfine anomaly\\

Isotope 2
	& The second isotope used.\\

Atomic state/ s-anomaly	& The atomic state for which the experimental hyperfine\\
                        & anomaly has been determined or the s-electron hfa.\\

$^{1}\Delta^{2}(\%)$
	& Hyperfine anomaly given in \%.\\

Reference
	& Original article where $^{1}\Delta^{2}(\%)$ or the experimental hyperfine\\
    & interactions constants is given.\\

\end{tabular}
\end{center}
\label{tableI}

\renewcommand{\arraystretch}{1.0}

\bigskip

\newpage




\setlength{\LTleft}{0pt}
\setlength{\LTright}{0pt}


\setlength{\tabcolsep}{0.5\tabcolsep}

\renewcommand{\arraystretch}{1.0}

\footnotesize 

\begin{longtable}{llllcr}
\caption{Experimental data of hyperfine anomaly values in atomic systems}\\
\mbox{Element} & \mbox{isotope 1} & \mbox{isotope 2} & Atomic state/ s-anomaly & $^{1}\Delta^{2}(\%)$ & Reference\\
\hline\\
\endfirsthead\\
\caption[]{(continued)}\\
\mbox{Element} & \mbox{isotope 1} & \mbox{isotope 2} & Atomic state/ s-anomaly & $^{1}\Delta^{2}(\%)$ & Reference\\
\hline\\
\endhead

Li & 6 & 7 & 2s $^{2}$S$_{1/2}$, s-anomaly & 0.00681(7) & \cite{BECKMANN1974}
\\
&  &  & 3s $^{2}$S$_{1/2}$, s-anomaly & 0.022(55) & \cite{Bushaw2003} \\
&  &  & 3p $^{2}$P$_{1/2}$ & -0.19(4) & \cite{Das2008} \\
\\
N & 14 & 15 & 2p$^{3}$ $^{4}$S$_{3/2}$ & 0.0999(4) & \cite{HOLLOWAY1962} \\
\\
Na & 23 & 24 & 3s $^{2}$S$_{1/2}$, s-anomaly & 0.0013(30) & \cite%
{BECKMANN1974}\cite{CHAN1966} \\
\\
Cl & 35 & 37 & 3p$^{5}$ $^{2}$P$_{3/2}$ & -0.00381(2) & \cite{Holloway1956}
\\
\\
K & 39 & 37 & 4s $^{2}$S$_{1/2}$, s-anomaly & -0.249(35) & \cite{PLATEN1971}
\\
& 39 & 40 & 4s $^{2}$S$_{1/2}$, s-anomaly & 0.466(19) & \cite{EISINGER1952}
\\
& 39 & 41 & 4s $^{2}$S$_{1/2}$, s-anomaly & -0.22936(14) & \cite%
{BECKMANN1974} \\
& 39 & 42 & 4s $^{2}$S$_{1/2}$, s-anomaly & 0.336(38) & \cite{CHAN1969} \\
\\
V & 50 & 51 & 3d$^{3}$4s$^{2}$ $^{4}$F$_{5/2}$ & 0.0007(10) & \cite{LUTZ1981}
\\
&  &  & 3d$^{4}$4s $^{6}$D$_{1/2}$ & 0.034(60) & \cite{Cochrane1998} \\
\\
Cu & 63 & 65 & 3d$^{10}$4s $^{2}$S$_{1/2}$ & 0.004861(9) & \cite{LUTZ1978}
\\
&  &  & 3d$^{9}$4s4p $^{4}$P$_{5/2}$ & 0.00340(11) & \cite{LUTZ1978}\cite%
{BLACHMAN1969} \\
&  &  & 3d$^{9}$4s4p $^{4}$P$_{9/2}$ & 0.00305(17) & \cite{LUTZ1978}\cite%
{BLACHMAN1969} \\
\\
Ga & 69 & 67 & 4p $^{2}$P$_{1/2}$ & -0.00050(7) & \cite{LUTZ1971} \\
&  &  & 4p $^{2}$P$_{3/2}$ & 0.00200(16) & \cite{LUTZ1971} \\
& 69 & 71 & 4p $^{2}$P$_{1/2}$ & 0.00063(6) & \cite{LUTZ1971} \\
&  &  & 4p $^{2}$P$_{3/2}$ & -0.00252(12) & \cite{LUTZ1971} \\
& 71 & 72 & 4p $^{2}$P$_{1/2}$ & 0.0043(6) & \cite{LUTZ1971} \\
&  &  & 4p $^{2}$P$_{3/2}$ & -0.0170(18) & \cite{LUTZ1971} \\
\\
Br & 79 & 81 & 4p$^{5}$ $^{2}$P$_{3/2}$ & -0.00003(4) & \cite{BROWN1966}\cite%
{LUTZ1970} \\
\\
Rb & 85 & 84 & 5s $^{2}$S$_{1/2}$, s-anomaly & -1.7(1.0) & \cite%
{ACKERMANN1973} \\
& 85 & 86 & 5s $^{2}$S$_{1/2}$, s-anomaly & 0.17(9) & \cite{BRASLAU1961} \\
& 85 & 87 & 5s $^{2}$S$_{1/2}$, s-anomaly & 0.35142(30) & \cite{BEDERSON1952}%
\cite{Arimondo1977} \\
&  &  & 6s $^{2}$S$_{1/2}$, s-anomaly & 0.36(2) & \cite{Galvan2007}\cite%
{Galvan2008} \\
&  &  & 7s $^{2}$S$_{1/2}$, s-anomaly & 0.32(2) & \cite{Chui2005} \\
&  &  & 5p $^{2}$P$_{1/2}$ & 0.673(7) & \cite{Das2008} \\
&  &  & 5p $^{2}$P$_{3/2}$ & 0.164(8) & \cite{Das2008} \\
\\
Mo & 95 & 97 & 4d$^{5}$5s $^{7}$S$_{3}$ & -0.0101(2) & \cite{BUTTGENBACH1974}
\\
\\
Ru & 99 & 101 & s-anomaly & -0.0173(1) & \cite{BUTTGENBACH1984} \\
\\
Ag & 107 & 103 & 4d$^{10}$5s $^{2}$S$_{1/2}$ & -3.4(1.7) & \cite%
{Wannberg1970} \\
& 107 & 108 & 4d$^{10}$5s $^{2}$S$_{1/2}$ & -2.6(7) & \cite{CUSSENS1969} \\
& 107 & 109 & 4d$^{10}$5s $^{2}$S$_{1/2}$ & -0.41274(29) & \cite{DAHMEN1967}
\\
& 107 & 109$^{m}$ & 4d$^{10}$5s $^{2}$S$_{1/2}$ & -3.8(4.1) & \cite%
{STINSON1971} \\
&  &  &  & -0.85(1.19) & \cite{STINSON1971}, $\mu _{I}$ from \cite{EDER1985}
\\
& 107 & 110 & 4d$^{10}$5s $^{2}$S$_{1/2}$ & -3.1(1.4) & \cite{CUSSENS1969}
\\
& 107 & 110$^{m}$ & 4d$^{10}$5s $^{2}$S$_{1/2}$ & -2.88(13) & \cite%
{SCHMELLI1967}\\
\\
Cd & 111 & 107 & 5s5p $^{3}$P$_{1}$ & -0.0958(8) & \cite{THADDEUS1963} \\
& 111 & 109 & 5s5p $^{3}$P$_{1}$ & -0.0912(7) & \cite{THADDEUS1963} \\
& 111 & 113 & 5s5p $^{3}$P$_{1}$ & -0.00023(40) & \cite{CHANEY1969} \\
& 111 & 113 & 5s5p $^{3}$P$_{2}$ & -0.00143(6) & \cite{FAUST1960} \\
& 111 & 113 & 5s6s $^{3}$S$_{1}$ & -0.01(4) & \cite{CHANTEPI1969} \\
& 111 & 113$^{m}$ & 5s5p $^{3}$P$_{1}$ & -0.0773(5) & \cite{CHANEY1969} \\
& 111 & 115 & 5s5p $^{3}$P$_{1}$ & 0.244(65) & \cite{CHANEY1969} \\
& 111 & 115$^{m}$ & 5s5p $^{3}$P$_{1}$ & -0.236(90) & \cite{CHANEY1969} \\
\\
In & 113 & 115 & 5p $^{2}$P$_{1/2}$ & 0.00075(13) & \cite{ECK1957} \\
&  &  & 5p $^{2}$P$_{3/2}$ & -0.00238(13) & \cite{ECK1957} \\
\\
Sn & 115 & 117 & 5p$^{2}$ $^{3}$P$_{1}$ & 0,0034(10) & \cite{CHILDS1965} \\
&  &  & 5p$^{2}$ $^{3}$P$_{2}$ & -0.0003(10) & \cite{CHILDS1965} \\
& 117 & 119 & 5p$^{2}$ $^{3}$P$_{1}$ & 0.0049(10) & \cite{CHILDS1965} \\
&  &  & 5p$^{2}$ $^{3}$P$_{2}$ & -0.0009(10) & \cite{CHILDS1965} \\
&  &  & 5p$^{2}$ $^{1}$D$_{2}$ & +0.0001(10) & \cite{CHILDS1971} \\
\\
Sb & 121 & 123 & 5p$^{3}$ $^{4}$S$_{3/2}$ & -0.323(9) & \cite{FERNANDO1960}
\\
\\
Cs & 133 & 131 & 6s $^{2}$S$_{1/2}$, s-anomaly & 0.48(5) & \cite{WORLEY1965}
\\
& 133 & 134 & 6s $^{2}$S$_{1/2}$, s-anomaly & 0.17(3) & \cite{STROKE1957} \\
& 133 & 134$^{m}$ & 6s $^{2}$S$_{1/2}$, s-anomaly & -1.38(3) & \cite%
{COHEN1962} \\
& 133 & 135 & 6s $^{2}$S$_{1/2}$, s-anomaly & 0.04(1) & \cite{STROKE1957} \\
& 133 & 137 & 6s $^{2}$S$_{1/2}$, s-anomaly & 0.0018(40) & \cite{STROKE1957}
\\
\\
Ba & 135 & 137 & 5d6s $^{3}$D$_{1}$ & -0.205(7) & \cite{GUSTAVSSON1979} \\
&  &  & 5d6s $^{3}$D$_{2}$ & -0.179(22) & \cite{GUSTAVSSON1979} \\
&  &  & 5d6s $^{3}$D$_{3}$ & -0.188(17) & \cite{GUSTAVSSON1979} \\
&  &  & 5d6s $^{1}$D$_{2}$ & -0.212(26) & \cite{SCHMELLING1974} \\
&  &  & Ba$^{+}$ 6s $^{2}$S$_{1/2}$, s-anomaly & -0.191(5) & \cite{Trapp2000}
\\
\\
La & 138 & 139 & 5d6p $^{3}$D$_{1}$ & -0.35(23) & \cite{Iimura2003}\cite%
{CHILDS1979} \\
\\
Nd & 143 & 145 & s-anomaly & 0.2034(63) & \cite{Persson2009} \\
\\
Eu & 151 & 145 & s-anomaly & -0.08(15) & \cite{Persson2009_1} \\
& 151 & 146 & s-anomaly & 0.12(50) & \cite{Persson2009_1} \\
& 151 & 147 & s-anomaly & -0.12(17) & \cite{Persson2009_1} \\
& 151 & 148 & s-anomaly & 0.08(31) & \cite{Persson2009_1} \\
& 151 & 149 & s-anomaly & -0.19(16) & \cite{Persson2009_1} \\
& 151 & 150 & s-anomaly & 0.08(28) & \cite{Persson2009_1} \\
& 151 & 152 & s-anomaly & 0.50(6) & \cite{Persson2009_1} \\
& 151 & 153 & s-anomaly & -0.64(3) & \cite{Brand1981}\\
\\
Gd & 155 & 157 & s-anomaly & 0.106(24) & \cite{Persson2009} \\
\\
Dy & 161 & 163 & 4f$^{10}$6s6p $^{5}$K$_{8}$ & 0.019(16) & \cite{CLARK1982}
\\
&  &  & 4f$^{10}$6s6p $^{5}$K$_{9}$ & 0.025(11) & \cite{CLARK1982} \\
&  &  & 4f$^{10}$6s6p $^{5}$I$_{8}$ & -0.116(19) & \cite{CLARK1982} \\
&  &  & 4f$^{10}$6s6p $^{5}$H$_{7}$ & -0.176(36) & \cite{CLARK1982} \\
\\
Yb & 171 & 173 & 6s6p $^{3}$P$_{1}$ & -0.386(5) & \cite{BUDICK1970} \\
&  &  & 4f$^{13}$5d6s$^{2}$ $^{3}$P$_{1}$ & 0.066(22) & \cite{BUDICK1970} \\
&  &  & Yb$^{+}$ 6s $^{2}$S$_{1/2}$, s-anomaly & -0.425(5) & \cite{Fisk1997}%
\cite{MUNCH1987} \\
\\
Lu & 175 & 176 & 5d6s$^{2}$ $^{2}$D$_{3/2}$ & 0.02(15) & \cite{BRENNER1985}
\\
&  &  & 5d6s$^{2}$ $^{2}$D$_{5/2}$ & 0.19(15) & \cite{BRENNER1985} \\
&  &  & 5d6s6p $^{4}$P$_{1/2}$ & 0.40(24) & \cite{Witte2002} \\
&  &  & 5d6s6p $^{4}$P$_{3/2}$ & 1.62(25) & \cite{Witte2002} \\
&  &  & 5d6s6p $^{4}$P$_{5/2}$ & 0.0(27) & \cite{Witte2002} \\
&  &  & 5d6s6p $^{4}$F$_{3,5,7/2}$, s-anomaly & 0.48(8) & \cite{KUHNERT1983}%
, $\mu _{I}$ from \cite{BRENNER1985} \\
&  &  & 6s$^{2}$8p $^{2}$P$_{1/2}$ & 1.84(90) & \cite{Witte2002} \\
&  &  & 6s$^{2}$8p $^{2}$P$_{3/2}$ & 0.55(22) & \cite{Witte2002} \\
& 175 & 177 & s-anomaly & -0.018(35) & \cite{BRENNER1985} \\
& 176 & 176$^{m}$ & s-anomaly & 0.48(8) & \cite{BRENNER1985} \\
\\
Re & 185 & 186 & 5d$^{5}$6s$^{2}$ $^{6}$S$_{5/2}$ & -1.36(17) & \cite%
{Armstrong1965}\cite{Buttgenbach1981} \\
& 185 & 187 & 5d$^{5}$6s$^{2}$ $^{6}$S$_{5/2}$ & 0.031(8) & \cite%
{Armstrong1965} \\
& 185 & 187 & s-anomaly & 0.027(5) & \cite{Burger1982} \\
& 185 & 188 & 5d$^{5}$6s$^{2}$ $^{6}$S$_{5/2}$ & -1.28(28) & \cite%
{Armstrong1965}\cite{Buttgenbach1981} \\
\\
Ir & 191 & 193 & s-anomaly & -0.64(7) & \cite{BUTTGENBACH1978-1} \\
\\
Au & 197 & 196 & 6s $^{2}$S$_{1/2}$, s-anomaly & 8.69(26) & \cite%
{SCHMELLI1970}\cite{EKSTROM1980} \\
& 197 & 198 & 6s $^{2}$S$_{1/2}$, s-anomaly & 8.53(8) & \cite{VANDENBO1967}
\cite{EKSTROM1980} \\
& 197 & 199 & 6s $^{2}$S$_{1/2}$, s-anomaly & 3.64(29) & \cite{VANDENBO1967}
\cite{EKSTROM1980} \\
\\
Hg & 199 & 193 & 6s6p $^{3}$P$_{1}$ & -0.61(3) & \cite{REIMANN1973} \\
& 199 & 193$^{m}$ & 6s6p $^{3}$P$_{1}$ & -1.0552(13) & \cite{REIMANN1973} \\
& 199 & 195 & 6s6p $^{3}$P$_{1}$ & -0.1470(9) & \cite{REIMANN1973} \\
& 199 & 195$^{m}$ & 6s6p $^{3}$P$_{1}$ & -1.038(3) & \cite{REIMANN1973} \\
& 199 & 197 & 6s6p $^{3}$P$_{1}$ & -0.0778(7) & \cite{REIMANN1973} \\
& 199 & 197$^{m}$ & 6s6p $^{3}$P$_{1}$ & -1.021(3) & \cite{REIMANN1973} \\
& 199 & 199$^{m}$ & 6s6p $^{3}$P$_{1}$ & --0.960(9) & \cite{REIMANN1973} \\
& 199 & 201 & 6s6p $^{3}$P$_{1}$ & -0.1467(6) & \cite{REIMANN1973} \\
& 199 &  & 6s6p $^{3}$P$_{2}$ & -0.15653(4) & \cite{REIMANN1973} \\
& 199 &  & Hg$^{+}$, 6s $^{2}$S$_{1/2}$, s-anomaly & -0.16257(5) & \cite
{Burt2009} \\
& 199 & 203 & 6s6p $^{3}$P$_{1}$ & -0.796(16) & \cite{REIMANN1973} \\
\\
Tl & 203 & 205 & 6p $^{2}$P$_{1/2}$ & 0.01035(15) & \cite{LURIO1956} \\
&  &  & 6p $^{2}$P$_{3/2}$ & -0.16258(10) & \cite{GOULD1956} \\
\\
Fr & 212 & 208 & 7s $^{2}$S$_{1/2}$ - 7p $^{2}$P$_{1/2}$ ($\Delta
_{s}-\Delta _{p}$) & -0.032(38) & \cite{Grossman1999} \\
& 212 & 209 & 7s $^{2}$S$_{1/2}$ - 7p $^{2}$P$_{1/2}$ ($\Delta _{s}-\Delta
_{p}$) & 0.339(31) & \cite{Grossman1999} \\
& 212 & 210 & 7s $^{2}$S$_{1/2}$ - 7p $^{2}$P$_{1/2}$ ($\Delta _{s}-\Delta
_{p}$) & -0.007(28) & \cite{Grossman1999} \\
& 212 & 211 & 7s $^{2}$S$_{1/2}$ - 7p $^{2}$P$_{1/2}$ ($\Delta _{s}-\Delta
_{p}$) & 0.331(34) & \cite{Grossman1999} \\
\\
U & 233 & 235 & 5f$^{3}$6d7s$^{2}$ $^{5}$L$_{6}$ & 0.84(31) & \cite%
{Gangrsky1997} \\
&  &  & 5f$^{3}$6d7s7p $^{7}$M$_{7}$ & 1.32(31) & \cite{Gangrsky1997} \\
&  &  & 5f$^{3}$6d7s7p $^{7}$L$_{6}$ & 1.19(89) & \cite{Gangrsky1997} \\
\end{longtable}

\newpage

\renewcommand{\arraystretch}{1.0}


\begin{thebibliography}{999}
                                                                                              %
\bibitem {otten}E.W. Otten, Treatise on Heavy-Ion Science\textit{\ }vol 8, ed.
D. Allan Bromely (Plenum, New York 1989) 515

\bibitem {billowes}J.Billowes and P. Campbell, J. Phys. G
\textbf{21}, (1995) 707.

\bibitem {Buttgenbach}S. B\"{u}ttgenbach, Hyperfine Int. \textbf{20}, (1984) 1

\bibitem {stroke}H.H. Stroke, V. Jaccarino, D.S. Edmonds, Jr and
R. Weiss, Phys. Rev. \textbf{105}, (1957) 590.

\bibitem {mosko}P.A. Moskowitz, C.H. Liu, G. Fulop and H.H. Stroke, Phys. Rev. \textbf{C4}, (1971) 620

\bibitem {Kopf}  H. Kopfermann, Kernmomente (Akademische Verlagsgesellschaft, Leipzig, 1940)

\bibitem {Bitter} F. Bitter, Phys. Rev. \textbf{76},150 (1949)

\bibitem {bohrweisskopf}A. Bohr and V.F. Weisskopf, Phys. Rev. \textbf{77},
(1950) 94

\bibitem {Breit}E. Rosenthal and G. Breit, Phys. Rev. \textbf{41},
(1932) 459

\bibitem {Crawford}M. Crawford and A. Schawlow, Phys. Rev.
\textbf{76}, (1949) 1310

\bibitem {Pallas}N.J. Ionesco-Pallas, Phys. Rev. \textbf{117}, (1960) 505

\bibitem {Rosenberg}H.J. Rosenberg and H.H. Stroke, Phys. Rev. A
\textbf{5}, (1972) 1992

\bibitem{Galvan}
Perez Galvan, A., Zhao,  Y., Orozco, L.A., Gomez E., Lange, A.D., Baumer,A. and
  Sprouse, G.D.: Phys. Lett. \textbf{655B}, 114 (2007)

\bibitem {Ref4}I. Lindgren and J. Morrison, Atomic Many-Body Theory
(Springer-Verlag, Berlin 1983)

\bibitem {Ref5}P.G.H. Sandars and J. Beck, Proc. R. Soc. London \textbf{A289},
97 (1965)

\bibitem {persson}J.R. Persson, Euro. Phys. J. \textbf{A2},(1998) 3


\bibitem {stone}N.J. Stone At. Data Nucl. Data Tables \textbf{90} (2005) 75–176


\end{thebibliography}

\footnotesize

\newpage


\begin{thebibliography}{99}


\bibitem{BECKMANN1974}
A. Beckmann, K.D. Boklen and D. Elke
\newblock {\em {Z. Phys.}}, {270}({3}):{173--186}, {1974}.

\bibitem{Bushaw2003}
B.A. Bushaw, W. Nortershauser, G. Ewald, A. Dax and G.W.F. Drake
\newblock {\em {Phys. Rev. Lett.}}, {91}({4}), {JUL 25} {2003}.

\bibitem{Das2008}
D. Das and V. Natarajan
\newblock {\em {J. Phys. B}},
  {41}({3}), {FEB 14} {2008}.

\bibitem{HOLLOWAY1962}
W.W. HOlloway, R. Novick, and E. Luscher.
\newblock {\em {Phys. Rev.}}, {126}({6}):{2109--\&}, {1962}.

\bibitem{CHAN1966}
Y.W. Chan, V.W. Cohen, M. Lipsicas and H.B. Silsbee
\newblock {\em {Phys. Rev.}}, {150}({3}):{933--\&}, {1966}.

\bibitem{Holloway1956}
J.H. King, J.H. Holloway and B.B. Aubrey
\newblock Hyperfine structure and nuclear magnetic octopole moment of the
  stable chlorine isotopes.
\newblock Technical Report NP-6012, Massachusetts institute of Technology,
  1956.
\newblock Quarterly progress report, Research laboratory of electronics.

\bibitem{PLATEN1971}
C.V. Platen, J. Bonn, U. Kopf, R. Neugart and E.W. Otten
\newblock {\em {Z. Phys.}}, {244}({1}):{44--\&}, {1971}.

\bibitem{EISINGER1952}
J.T. Eisinger, B. Bederson and B.T. Feld
\newblock {\em {Phys. Rev.}}, {86}({1}):{73--81}, {1952}.

\bibitem{CHAN1969}
Y.W. Chan, V.W. Cohen and H.B. Silsbee
\newblock {\em {Phys. Rev.}}, {184}({4}):{1102--\&}, {1969}.

\bibitem{LUTZ1981}
O. Lutz, W. Messner, K.R. Mohn and P. Kroneck
\newblock {\em {Z. Phys. A}},
  {300}({2-3}):{111--114}, {1981}.

\bibitem{Cochrane1998}
E.C.A Cochrane, D.M. Benton, D.H. Forest and J.A.R. Griffith
\newblock {\em {J. Phys. B}},
  {31}({10}):{2203--2213}, {MAY 28} {1998}.

\bibitem{LUTZ1978}
O. Lutz, H. Oehler and P. Kroneck
\newblock {\em {Z. Phys. A}},
  {288}({1}):{17--21}, {1978}.

\bibitem{BLACHMAN1969}
A.G. Blachman, D.A. Landman and A. Lurio
\newblock {\em {Phys. Rev.}}, {181}({1}):{70--\&}, {1969}.

\bibitem{LUTZ1971}
O. Lutz, A. Nolle and A. Uhl
\newblock {\em {Z. Phys.}}, {248}({2}):{159--\&}, {1971}.

\bibitem{BROWN1966}
H.H. Brown and J.G. King
\newblock {\em {Phys. Rev.}}, {142}({1}):{53--\&}, {1966}.

\bibitem{LUTZ1970}
O. Lutz
\newblock {\em {Phys. Lett. A}}, {A 31}({7}):{384--\&}, {1970}.

\bibitem{ACKERMANN1973}
F. Ackermann, I. Platz and G. zu Pulitz
\newblock {\em {Z. Phys.}}, {260}({2}):{87--110}, {1973}.

\bibitem{BRASLAU1961}
N. Braslau, G.O. Brink and J.M. Khan
\newblock {\em {Phys. Rev.}}, {123}({5}):{1801--\&}, {1961}.

\bibitem{BEDERSON1952}
B. Bederson and V. Jaccarino
\newblock {\em {Phys. Rev.}}, {87}({1}):{228--229}, {1952}.

\bibitem{Arimondo1977}
E. Arimondo, M. Inguscio and P. Violino
\newblock {\em Rev. Mod. Phys.}, 49(1):31--75, Jan 1977.

\bibitem{Galvan2007}
A. Perez Galvan, Y. Zhao, L.A. Orozco, E. Gomez, A.D. Lange, F. Baumer and
  G.D. Sprouse
\newblock {\em {Phys. Lett. B}}, {655}({3-4}):{114--118}, {NOV 1} {2007}.

\bibitem{Galvan2008}
A. Perez Galvan, Y. Zhao and L.A. Orozco
\newblock {\em {Phys. Rev. A}}, {78}({1}), {JUL} {2008}.

\bibitem{Chui2005}
H.C. Chui, M.S. Ko, Y.W. Liu, J.T. Shy, J.L. Peng and H. Ahn
\newblock {\em {Opt. Lett.}}, {30}({8}):{842--844}, {APR 15} {2005}.

\bibitem{BUTTGENBACH1974}
S. Buttgenbach, M. Herschel, G. Meisel, E. Schrodl, W. Witte and W.J. Childs
\newblock {\em {Z. Phys.}}, {266}({4}):{271--274}, {1974}.

\bibitem{BUTTGENBACH1984}
S. Buttgenbach
\newblock {\em {Hyperfine Int.}}, {20}({1}):{1--64}, {1984}.

\bibitem{Wannberg1970}
B. Wannberg, J.O. Jonsson,and L. Sanner
\newblock {\em Phys. Scr.}, 1(5-6):238--242, 1970.

\bibitem{CUSSENS1969}
C.J. Cussens, G.K. Rochester and K.F. Smith
\newblock {\em {J. Phys. A }}, {2}({6}):{658--\&},
  {1969}.

\bibitem{DAHMEN1967}
H. Dahmen and S. Penselin
\newblock {\em {Z. Phys.}}, {200}({4}):{456--\&}, {1967}.

\bibitem{STINSON1971}
G.M. Stinson, A.R. Pierce, J.C. Waddington and R.G. Summersgate
\newblock {\em {Can. J. Phys.}}, {49}({7}):{906--\&}, {1971}.

\bibitem{EDER1985}
R. Eder, E. Hagn and E. Zech
\newblock {\em {Phys. Rev. C}}, {31}({1}):{190--196}, {1985}.

\bibitem{SCHMELLI1967}
S.G. Schmelling, V.J. Ehlers and H.A. Shugart
\newblock {\em {Phys. Rev.}}, {154}({4}):{1142--\&}, {1967}.

\bibitem{THADDEUS1963}
P. Thaddeus and M.N. Mcdermott
\newblock {\em {Phys. Rev.}}, {132}({3}):{1186--\&}, {1963}.

\bibitem{CHANEY1969}
R.L. Chaney and M.N. Mcdermott
\newblock {\em {Phys. Lett. A}}, {A 29}({2}):{103--\&}, {1969}.

\bibitem{FAUST1960}
W. Faust, M. Mcdermott and W. Lichten
\newblock {\em {Phys. Rev.}}, {120}({2}):{469}, {1960}.

\bibitem{CHANTEPI1969}
M. Chantepi
\newblock {\em {Comptes Rendus Hebdomadaires Des Seances De L ACademie Des
  Sciences Serie B}}, {269}({12}):{522--\&}, {1969}.

\bibitem{ECK1957}
T.G. Eck, A. Lurio and P. Kusch
\newblock {\em {Phys. Rev.}}, {106}({5}):{954--957}, {1957}.

\bibitem{CHILDS1965}
W.J. Childs and L.S. Goodman
\newblock {\em {Phys. Rev.}}, {137}({1A}):{A35--\&}, {1965}.

\bibitem{CHILDS1971}
W.J. Childs
\newblock {\em {Phys. Rev. A}}, {4}({2}):{439--\&}, {1971}.

\bibitem{FERNANDO1960}
P.C.B Fernando, G.K. Rochester, I.J. Spalding and K.F. Smith
\newblock {\em {Phil. Mag.}}, {5}({60}):{1291--1298}, {1960}.

\bibitem{WORLEY1965}
R.D. Worley, V.J. Ehlers, W.A. Nierenberg and H.A. Shugart
\newblock {\em {Phys. Rev.}}, {140}({6B}):{1483--\&}, {1965}.

\bibitem{STROKE1957}
H.H. Stroke, V. Jaccarino, D.S. Edmonds and R. Weiss
\newblock {\em {Phys. Rev.}}, {105}({2}):{590--603}, {1957}.

\bibitem{COHEN1962}
V.W. Cohen, T. Moran and S. Penselin
\newblock {\em {Phys. Rev.}}, {127}({2}):{517--\&}, {1962}.

\bibitem{GUSTAVSSON1979}
M. Gustavsson, G. Olsson and A. Rosen
\newblock {\em {Z. Phys. A}},
  {290}({3}):{231--243}, {1979}.

\bibitem{SCHMELLING1974}
S.G. Schmelling
\newblock {\em {Phys. Rev. A}}, {9}({3}):{1097--1102}, {1974}.

\bibitem{Trapp2000}
S. Trapp, G. Marx, G. Tommaseo, A. Klaas, A. Drakoudis, G. Revalde, G. Szawiola and
  G. Werth
\newblock {\em {hyperfine int.}}, {127}({1-4}):{57--64}, {2000}.

\bibitem{Iimura2003}
H. Iimura, M. Koizumi, M. Miyabe, M. Oba, T. Shibata, N. Shinohara, Y. Ishida,
  T. Horiguchi and H.A. Schuessler
\newblock {\em {Phys. Rev. C}}, {68}({5}), {NOV} {2003}.

\bibitem{CHILDS1979}
W.J. Childs and L.S. Goodman
\newblock {\em {Phys. Rev. A}}, {20}({5}):{1922--1926}, {1979}.

\bibitem{Persson2009}
J.R. Persson
\newblock arXiv:0904.0516 {2009}.

\bibitem{Persson2009_1}
J.R. Persson
\newblock arXiv:0904.4828 {2009}.

\bibitem{Brand1981}
H. Brand, V. Pfeufer, and A. Steudel
\newblock {\em {Z. Phys. A}},
  {302}({4}):{291--298}, {1981}.

\bibitem{CLARK1982}
D.L. Clark and G.W. Greenlees
\newblock {\em {Phys. Rev. C}}, {26}({4}):{1636--1648}, {1982}.

\bibitem{BUDICK1970}
B. Budick and J. Snir
\newblock {\em {Phys. Rev. A}}, {1}({3}):{545--\&}, {1970}.

\bibitem{Fisk1997}
P.T.H Fisk, M.J. Sellars, M.A. Lawn and C. Coles
\newblock {\em {IEEE Transactions on Ultrasonics Ferroelectrics and Frequency
  Control}}, {44}({2}):{344--354}, {MAR} {1997}.

\bibitem{MUNCH1987}
A. Munch, M. Berkler, C. Gerz, D. Wilsdorf and G. Werth
\newblock {\em {Phys. Rev. A}}, {35}({10}):{4147--4150}, {MAY 15} {1987}.

\bibitem{BRENNER1985}
T. Brenner, S. Buttgenbach, W. Rupprecht and F. Traber
\newblock {\em {NucL. Phys. A}}, {440}({3}):{407--423}, {1985}.

\bibitem{Witte2002}
S. Witte, EJ. van Duijn, R. Zinkstok and W. Hogervorst
\newblock {\em {Eur. Phys. J. D}}, {20}({2}):{159--164}, {AUG}
  {2002}.

\bibitem{KUHNERT1983}
A. Kuhnert, A. Nunnemann and D. Zimmermann
\newblock {\em {J. Phys. B}},
  {16}({23}):{4299--4303}, {1983}.

\bibitem{Armstrong1965}
L. Armstrong and R. Marrus
\newblock {\em {Phys. Rev.}}, {138}({2B}):{B310--\&}, {1965}.

\bibitem{Buttgenbach1981}
S. Buttgenbach, R. Dicke, G. Golz and F. Traber
\newblock {\em {Z. Phys. A}},
  {302}({4}):{281--290}, {1981}.

\bibitem{Burger1982}
K.H. Burger, B. Burghardt, S. Buttgenbach, R. Harzer, H. Hoeffgen, G. Meisel and
  F. Traber
\newblock {\em {Z. Phys. A}},
  {307}({3}):{201--209}, {1982}.

\bibitem{BUTTGENBACH1978-1}
S. Buttgenbach, R. Dicke, H. Gebauer, R. Kuhnen and F. Traber
\newblock {\em {Z. Phys. A}},
  {286}({4}):{333--340}, {1978}.

\bibitem{SCHMELLI1970}
S.G. Schmelling, V.J. Ehlers and H.A. Shugart
\newblock {\em {Phys. Rev. C}}, {2}({1}):{225--\&}, {1970}.

\bibitem{EKSTROM1980}
C. Ekstrom, L. Robertsson, S. Ingelman, G. Wannberg and I. Ragnarsson
\newblock {\em {Nucl. Phys. A}}, {348}({1}):{25--44}, {1980}.

\bibitem{VANDENBO1967}
P.A. van den Bout., V.J. Ehlers, W.A. Nierenberg and H.A. Shugart
\newblock {\em {Phys. Rev.}}, {158}({4}):{1078--\&}, {1967}.

\bibitem{REIMANN1973}
R.J. Reimann and M.N. Mcdermott
\newblock {\em {Phys. Rev. C}}, {7}({5}):{2065--2079}, {1973}.

\bibitem{Burt2009}
E.A. Burt, S. Taghavi-Larigani, and R.L. Tjoelker
\newblock {\em {Phys. Rev. A}}, {79}({6}), {JUN} {2009}.

\bibitem{LURIO1956}
A. Lurio and A.G. Prodell
\newblock {\em {Phys. Rev.}}, {101}({1}):{79--83}, {1956}.

\bibitem{GOULD1956}
G. Gould
\newblock {\em {Phys. Rev.}}, {101}({6}):{1828--1829}, {1956}.

\bibitem{Grossman1999}
J.S. Grossman, L.A. Orozco, M.R. Pearson, J.E. Simsarian, G.D. Sprouse and W.Z. Zhao
\newblock {\em {Phys. Rev. Lett.}}, {83}({5}):{935--938}, {AUG 2}
  {1999}.

\bibitem{Gangrsky1997}
Y.P. Gangrsky, B.K. Kuldjanov, K.P. Marinova, B.N. Markov and S.G. Zemlyanoi
\newblock {\em {Z. Phys. D}},
  {42}({1}):{1--4}, {OCT} {1997}.


\end{thebibliography}
\end{document}